\documentclass[conference]{IEEEtran}
\IEEEoverridecommandlockouts
\usepackage{cite}
\usepackage{amsmath,amssymb,amsfonts}
\usepackage{algorithmic}
\usepackage{graphicx}
\usepackage{textcomp}
\usepackage{xcolor}
\usepackage{multirow}
\usepackage{url}
\usepackage{listings}
\usepackage{xcolor}
\usepackage{float}
\usepackage{makecell}
\def\BibTeX{{\rm B\kern-.05em{\sc i\kern-.025em b}\kern-.08em
    T\kern-.1667em\lower.7ex\hbox{E}\kern-.125emX}}
\renewcommand\IEEEkeywordsname{Keywords}
\begin{document}

\title{Performance Evaluation of Delay Tolerant Network Protocols to Improve Nepal Earthquake Rescue Communications \\
}

\author{\IEEEauthorblockN{Xiaofei Liu}
\IEEEauthorblockA{\textit{School of Computer Science} \\
\textit{University of Nottingham}\\
Nottingham, United Kingdom \\
psxxl24@nottingham.ac.uk}
\and
\IEEEauthorblockN{Milena Radenkovic}
\IEEEauthorblockA{\textit{School of Computer Science} \\
\textit{University of Nottingham}\\
Nottingham, United Kingdom \\
milena.radenkovic@nottingham.ac.uk}
}

\maketitle

\begin{abstract}
In the fields of disaster rescue and communication in extreme environments, Delay Tolerant Network (DTN) has become an important technology due to its "store-carry-forward" mechanism. Selecting the appropriate routing strategy is of crucial significance for improving the success rate of distress message transmission and reducing delays in material dispatch. 

We design a pseudo realistic use case of Nepal Kathmandu earthquake rescue based on dynamically changing population distribution model and characteristics of rescue activities in the initial rescue efforts in Nepal Kathmandu earthquakes to conducted the multi criteria two benchmark routing protocols performance analysis in the face of different buffer sizes of the rescue team nodes. We identify multiple real world node groups, including affected residents, rescue teams, drones and ground vehicles and communication models are established according to the movement behaviors of these groups. We analyze the communication of distress messages between edge nodes to obtain performance metrics such as delivered probability, average delay, hop count, and buffer time. By analyzing the multi layer complex data and protocols differences, the research results show the effectiveness of distributed DTN communication methods in the Nepal earthquake rescue use case, reveal existence of trade-offs between transmission reliability and resource utilization of different routing protocols in disaster communication environment and provide a basis for the design of next-generation emergency communication services based on edge nodes.
\end{abstract}


\section{Introduction}
This research adopted the 2015 Nepal earthquake as a sample to study. It has been strongly destructive and devastating in April 2015 with a magnitude score of 7.8, which caused the deaths of over 8900 and left about 20000 people injured\cite{b1}. The country's communication infrastructure has been severely affected. It is estimated that the total losses of telecommunications operators exceed 45.5 million US dollars\cite{b2}.This exemplifies that shortly after the earthquake, the local system of communication and transportation was disturbed to a magnified scale, which somehow affected the rescue operations.

When nodes are scattered and end-to-end connections are difficult to establish, the smartphones of survivors, the tablets of rescue teams, the vehicles of volunteers, and drones can all serve as nodes for data transmission and relaying\cite{b3}\cite{b4}. The adoption of Delay Tolerant Network (DTN) and opportunistic networks is an effective solution to the communication problems in rescue operations.In order to deeply explore how to build an efficient mobile ad hoc network in extreme environments, this study selects the 2015 Nepal earthquake as an analysis case, and by restoring its rescue communications, compares the performance of Epidemic and Spray-and-Wait routing protocols in the rescue communications.

\section{Relative Work}
Mobile ad hoc networks(MANETs) are an important type of wireless network, and one of their significant features is decentralization. Due to the continuous occurrence of such changes in mobile ad hoc networks, the topology of the network also keeps changing, which is referred to as "dynamic topology"\cite{b5}. VANETs have distinct characteristics such as high-speed movement of nodes, frequent changes in network topology, short connection time of communication links, and limitations imposed by road layouts\cite{b6}.In recent years, VANETs have been gradually evolving towards the Internet of Vehicles. Combining cloud computing, edge computing (as well as 5G/6G communication technologies), the modern vehicle networking architecture through heterogeneous network integration, it enables real-time interaction between vehicles and the cloud\cite{b7}. 

Delay/Disruption Tolerant Network(DTN) is a new type of network architecture designed for challenging network environments. Typical applications include wireless networks in remote areas, military networks, disaster emergency communication, and vehicular networks, mobile social networks, etc. In these scenarios, there often is no stable end-to-end path, and the round-trip delay can be measured in minutes or even hours. DTN uses a storage-carry-forward mechanism to cache messages at intermediate nodes for a long time and forwards them hop by hop when there is an opportunity to reach the target node, thus enabling final delivery even in the absence of continuous connectivity\cite{b8}.

Epidemic Routing is fundamental and classic type of routing protocol in the DTN field. It was proposed by Amin Vahdat and David Becker in an experimental technology report\cite{b9}. When two nodes meet, both parties simultaneously synchronize the messages that the other does not have, causing the copies of the messages in the network to spread exponentially until they reach the destination node or the system clears the messages\cite{b10}. The Spray and Wait protocol aims to reduce the resource consumption of flood-based routing. It maintains a high delivery success rate while limiting the number of message replicas\cite{b12}. By comparing these two classic protocols, this study analyzed the trade-off between transmission efficiency and resource occupation in the DTN rescue scenario.

The new generation of DTN routing protocols has an advantage in rescue use cases due to its computation of mobility, context cache awareness and contact history. CafRepCache extends CafRep by introducing context-aware caching and message eviction based on movement patterns, social proximity, and contact history\cite{b16}, potentially improving buffer efficiency and reducing premature message discarding in disaster environments. CognitiveCache enables adaptive learning-based caching strategies and collaboration among edge nodes, supporting dynamic cache placement under uncertain connectivity\cite{b17}. The unified interdependent multi-layer spatio-temporal network modeling framework proposed by V. S. H. Huynh and M. Radenkovic integrates space, time, and multi-layer structures into a single analytical system\cite{b18}, indicating future caching strategies may incorporate layered spatio-temporal prediction with dynamic adaptation.

\section{Nepal Rescue Use Case and Edge Communication Architecture Deployment}

This experiment designed a multi-layered communication architecture during the Nepal earthquake: the collapse of base stations, power grid failures, and the disruption of communication networks. The temporary command center and communication base stations had not yet been established, but rescue teams had already begun rescue operations. This section establishes a near-realistic framework for evaluating distributed rescue services by identifying different node groups and their unique movement patterns in earthquake rescue.

\subsection{Operation and Spatiotemporal Environment Design}

Table \ref{t1} shows the critical spatial and temporal specifications required for distributed emergency communications following the 2015 Nepal earthquake. The total duration is set to 12 hours (43,200 seconds).The first 72 hours after an earthquake are known as the golden rescue period. The longer injured people are buried in the rubble, the higher their risk of dying from compression and dehydration\cite{b23}\cite{b24}. A 12-hour case can fully cover the most critical daytime or rescue shift cycle after an earthquake. The use case area is set to 4500*3400, representing a urban-rural disaster landscape where infrastructure has collapsed. The warm-up time is set to 1000 seconds, which is sufficient for victims, rescue teams, and vehicles to disperse to various locations on the map according to their respective movement models.  This framework ensures that the evaluation of routing protocols is conducted within a stabilized, pseudo-realistic environment of node distribution.

\begin{table}[htbp]
    \centering
    \caption{Spatiotemporal Environment Design in Nepal Earthquake Use Case}
    \begin{tabular}{cc}
    \hline
        \textbf{Environmental Configuration} & \textbf{Value} \\ \hline
        EndTime & 43200 \\ 
        WorldSize & 4500,3400 \\ 
        WarmUp & 1000 \\ \hline
    \end{tabular}
    \label{t1}
\end{table}

\subsection{Edge Node Identification and Communication Interface Design}

\begin{figure}[htbp]
    \centering
    \includegraphics[width=1\linewidth]{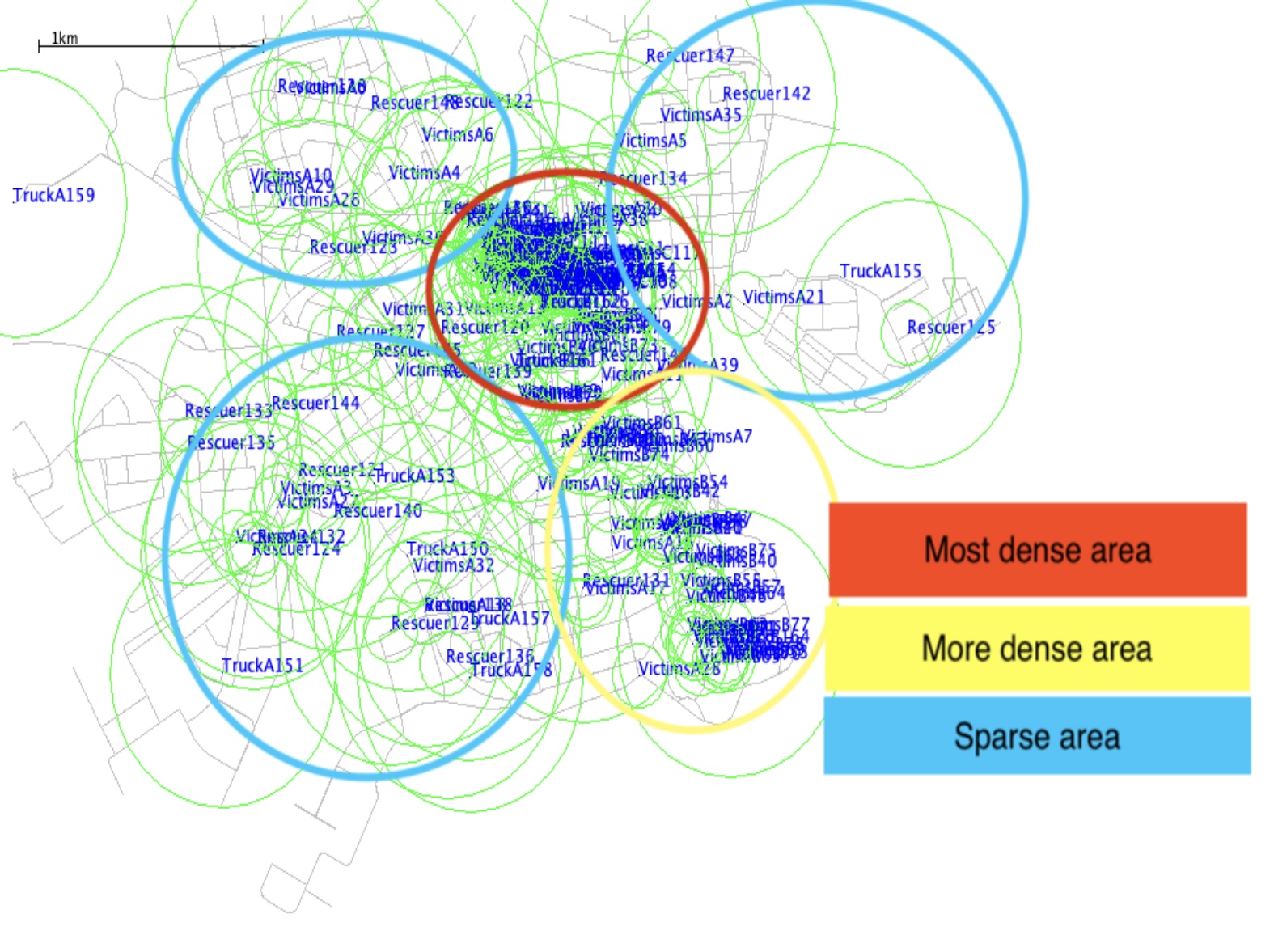}
    \caption{Screenshot illustration in Nepal Earthquake Use Case}
    \label{fig:6}
\end{figure}

\begin{figure}[htbp]
    \centering
    \includegraphics[width=0.8\linewidth]{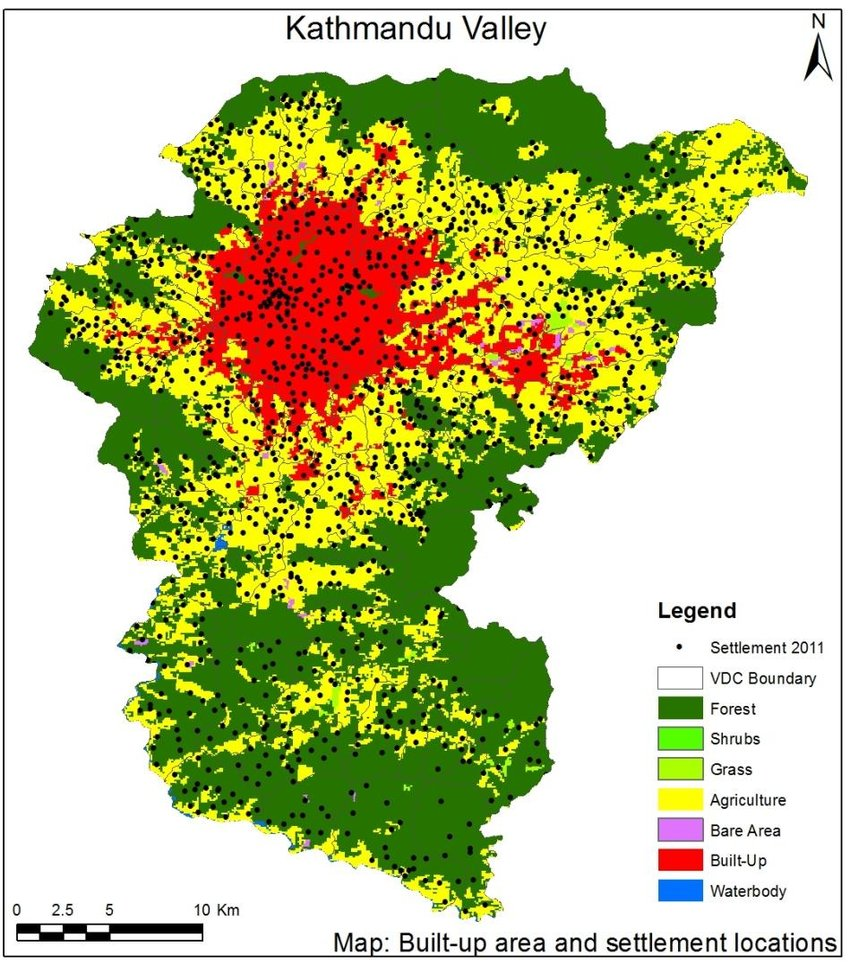}
    \caption{Location of built-up area and settlement along with other land cover areas in Kathmandu, Nepal\cite{b14}}
    \label{fig:7}
\end{figure}

To ensure a pseudo-realistic representation of the post-earthquake environment,we identify four distinct categories of edge nodes are identified based on their roles in the rescue operation and their underlying hardware capabilities. In Table \ref{t2} and Fig.\ref{fig:6}, The Victims nodes are the most numerous edge nodes at the scene. According to the population density areas of Kathmandu, they are divided into three groups: sparsely populated areas (VictimsA), densely populated areas (VictimsB), and highly populated built-up areas (VictimsC). The energy and storage capacity of these nodes are very limited, and they only rely on low-power Bluetooth interfaces to propagate distress signals. Dhonju et al.'s study revealed that Kathmandu\cite{b14}, Nepal's settlements, are not equally distributed; the crowded spaces are in the built-up and agricultural areas, while the forest zones are sparsely populated. According to Fig.\ref{fig:7}, it portrays the non-uniform population density of Kathmandu, which is in perfect accord with our node assembly in this experiment.

The rescue teams were evenly distributed across the map, demonstrating that they had not yet identified the disaster-stricken hotspots when they began their rescue operations immediately after the earthquake. Truck and drone nodes were divided into Group A and Group B. Group A had more nodes and covered all areas on the map, while Group B had fewer nodes and was specifically deployed in more densely populated areas. This reflects the importance of road network density and the speed at which aerial units can identify disaster areas.

\begin{table*}[htbp]
    \centering
    \caption{Communication Group Architecture Description in Nepal Earthquake Use Case}
    \begin{tabular}{cccccccc}
    \hline
        \textbf{Group} & \textbf{Name} & \textbf{Quantities} & \textbf{interface} & \textbf{okMaps} & \textbf{Speed} & \textbf{BufferSize} & \textbf{Message TTL} \\ \hline
        1 & VictimsA & 40 & Bluetooth & roads.wkt & 0.5,1.2 & 5M & 1200 \\ 
        2 & VictimsB & 40 & Bluetooth & pedestrian\_paths.wkt & 0.1,0.3 & 5M & 1200 \\ 
        3 & VictimsC & 40 & Bluetooth & shops.wkt & 0.1,0.3 & 5M & 1200 \\ 
        4 & Rescuer & 30 & Bluetooth,HighSpeed & roads.wkt & 1.2,2.8 & 10M/50M/100M & 2400 \\ 
        5 & TruckA & 10 & Bluetooth,HighSpeed & roads.wkt & 5,10 & 200M & 3600 \\ 
        6 & TruckB & 5 & Bluetooth,HighSpeed & pedestrian\_paths.wkt, shops.wkt & 5,10 & 200M & 3600 \\ 
        7 & DroneA & 8 & Bluetooth,HighSpeed & roads.wkt & 5,15 & 100M & 2400 \\ 
        8 & DroneB & 4 & BluetoothHighSpeed & pedestrian\_paths.wkt, shops.wkt & 5,15 & 100M & 2400 \\ \hline
    \end{tabular}
    \label{t2}
\end{table*}

\begin{table}[H]
    \centering
    \caption{Communication interface differences in Nepal Earthquake Use Case}
    \begin{tabular}{ccc}
    \hline
        \textbf{Interface} & \textbf{Transmit Speed} & \textbf{Transmit Range} \\ \hline
        btInterface & 2M & 120 \\ 
        highSpeedInterface & 10M & 500 \\ \hline
    \end{tabular}
    \label{tx}
\end{table}

Table \ref{tx} shows two interface used in Nepal earthquake rescue use case. btInterface is a Bluetooth interface with a transmission speed of 2Mb/s and a range of 120 meters. highSpeedInterface is a high-speed interface with a transmission speed of 10Mb/s and a range of 500 meters. Victims uses only the Bluetooth interface, while rescue teams, trucks, and drones use both Bluetooth and the high-speed interface. The design of the two interfaces is primarily to reflect the heterogeneity of equipment and communication capabilities at different nodes in actual earthquake rescue operations.

\subsection{Message Broadcast Logic and Routing Strategies}
In the early stage of earthquake rescue, the successful transmission of the SOS signals sent by the victims to the rescue team is of vital importance. This experiment conducted an exercise to assess whether the distress signals sent by the victims can be received by the rescue team in a timely manner. Table \ref{t3} presents the event design of the SOS signals sent by the victims. All the victims would send an SOS signal every 60 to 120 seconds, which shows the real situation where the rescue team receive a large number of distress signals in the early stage of earthquake rescue. Moreover, in the early stage of earthquake rescue, the rescue team cannot promptly assess the situation of the victims, so this use case adopted the method where all the victims randomly sent SOS signals to a certain rescue team, effectively covering the short, medium, and long-distance connections that occurs in the rescue communication.

In terms of routing strategies, for Epidemic Routing, nodes synchronize upon contact to maximize the reachability of distress signals. Spray-and-Wait strategy is configured with 16 copies, mainly considering the situation with a large number of nodes. At the same time, the binary mode was enabled to increase the diffusion speed of the replicas in the communications, which is suitable for earthquake disaster use case with limited resources and limited contact opportunities.

\begin{table}[htbp]
    \centering
    \caption{Victims SOS Events description in Nepal Earthquake Use Case}
    \begin{tabular}{cccccc}
    \hline
        \textbf{Host} & \textbf{toHost} & \textbf{Name} & \textbf{Interval} & \textbf{DataSize} \\ \hline
        Victims (A, B, C) & Rescure & SOS & 60,120 & 500k,1M \\ \hline
    \end{tabular}
    \label{t3}
\end{table}

\section{Results and Evaluation}
\subsection{Multi-criteria comparison of two protocols performance}

Table \ref{t4} shows the generation and delivery of messages under different protocols. Since the same seed was used, the number of messages created in different experiments was the same. In this use case, 484 messages were created in 12 hours. The data were all collected with the Rescuer node buffersize set to 50M in this section.

\begin{table}[htbp]
    \centering
    \caption{Delivery Performance Comparison Between Epidemic and Spray and Wait in Nepal Earthquake Use Case}
    \begin{tabular}{ccc}
    \hline
        \textbf{Massage status} & \textbf{Epidemic} & \textbf{Spray and Wait} \\ \hline
        Created & 484 & 484 \\ 
        Delivered & 75 & 458 \\ 
        Delivered Probability & 0.1550 & 0.9463 \\ \hline
    \end{tabular}
    \label{t4}
\end{table}

\begin{figure}[htbp]
    \centering
    \includegraphics[width=1\linewidth]{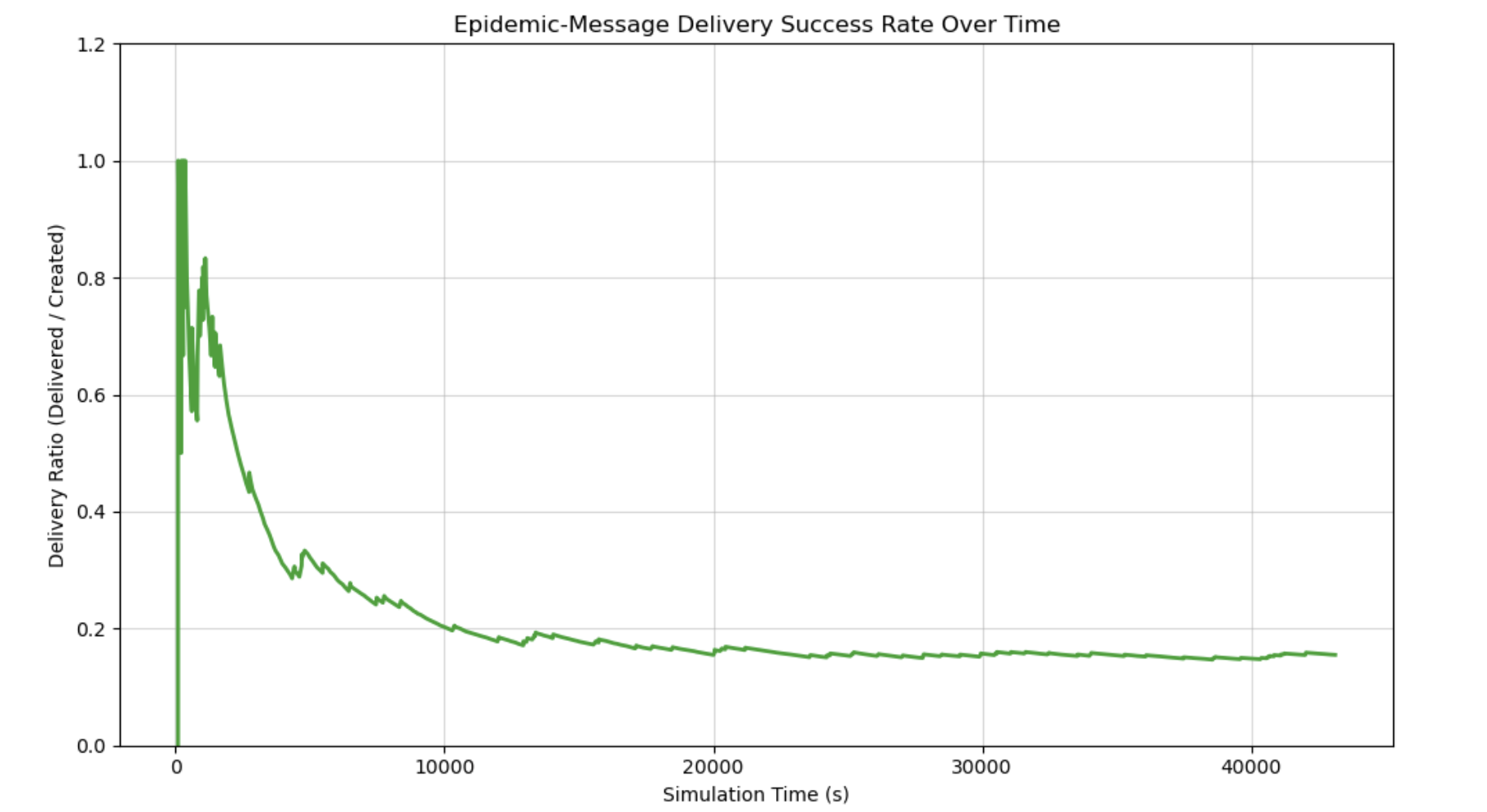}
    \caption{Message Delivery Success Rate Over Time of the Epidemic Protocol in Nepal Earthquake Use Case}
    \label{fig:9}
\end{figure}

\begin{figure}[htbp]
    \centering
    \includegraphics[width=1\linewidth]{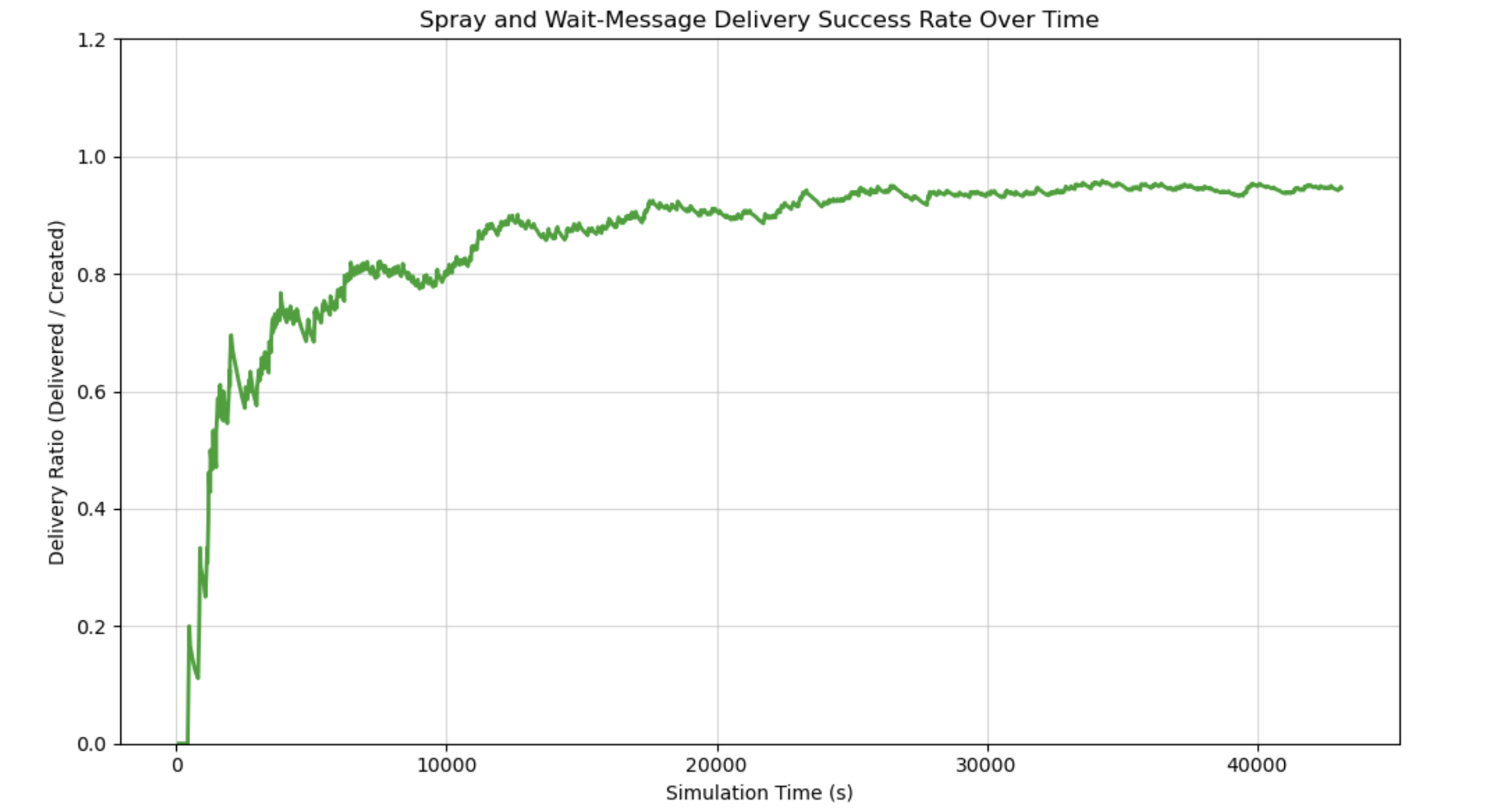}
    \caption{Message Delivery Success Rate Over Time of the Spray and Wait Protocol in Nepal Earthquake Use Case}
    \label{fig:10}
\end{figure}

Delivery probability is a key indication of the performance of protocols, which is summarized in Table \ref{t4}. In the case of the Epidemic protocol, 75 out of 484 messages were delivered, resulting in a delivery probability that stands at 15.50\%. This can be interpreted as very high network congestion, which eventually leads to network collapses. A glance at Fig.\ref{fig:9} depicts the gradual success rate of the Epidemic protocol in delivering messages within a given timeline. At the outset of the simulator run, minimal messages were generated. Most of these were short-distance direct deliveries, and therefore, the delivery rate appeared quite high, though it was just temporary. But, as time goes on, there are more messages generated. The problem with the Epidemic method was that the simulation started with trying to copy the messages to every other node it came across. After a while, the nodes' buffers were completely filled, and when the buffers overflowed, anything beyond the SOS messages or the messages received from the refugees was discarded, leading to a situation of clear network congestion and collapse and a rapid drop in the delivery rate. The delivery rate subsequently remained low. Instead, the Spray and Wait protocol delivered 458 out of 484 messages, with a 94.63\% delivery rate. A comparison of the delivery success rates of the Spray and Wait protocol over a period is provided in Fig.\ref{fig:10}. Evidently, the delivery success rate may be seen to have increased, indicating the fact that the increase in the number of message replicas eventually allowed discrete messages to get to the destination field over a sequence of transmission hops. This demonstrates that Spray and Wait effectively monitors the network traffic, preventing undue network congestion and collapse by controlling the number of replicas.

\begin{figure}[htbp]
    \centering
    \includegraphics[width=1.0\linewidth]{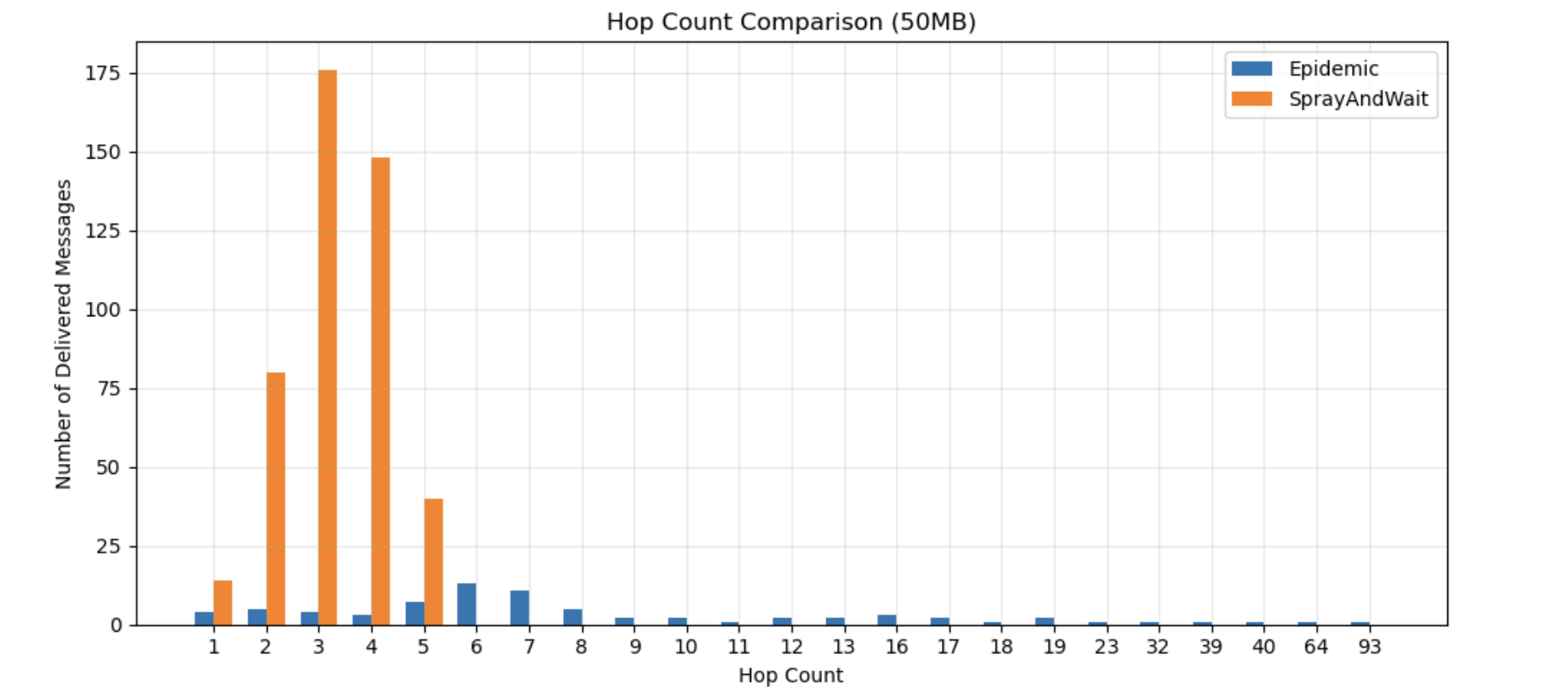}
    \caption{Hop Count Comparison Between Epidemic and Spray and Wait in Nepal Earthquake Use Case}
    \label{fig:11}
\end{figure}

As for the destinations of data foward, two protocols reveal a considerable disparity, as presented in Table \ref{t5}. Propagation of the Epidemic protocol, like being on a flood, necessitated an over-3-million-creation bundle of messages, whereas the overhead ratio was at 45,794.5733, implying that such countless invalid sends (over 45,000) had to be made across the network just to send one intended message. Therefore, in terms of bandwidth resources and device battery time, this was highly wasting specifically in earthquake communications. The Spray and Wait technique had the overhead ratio of 15.4, denoting a comparatively healthier level of resource occupancy.

\begin{table}[htbp]
    \centering
    \caption{Massage Replicate Comparison Between Epidemic and Spray and Wait in Nepal Earthquake Use Case}
    \begin{tabular}{ccc}
    \hline
        \textbf{Massage status} & \textbf{Epidemic} & \textbf{Spray and Wait} \\ \hline
        Started & 3,441,993 & 7,539 \\ 
        Relayed & 3,434,668 & 7,532 \\ 
        Dropped & 3,430,499 & 4,900 \\ 
        Overhead Ratio & 45,794.5733 & 15.4454 \\ \hline
    \end{tabular}
    \label{t5}
\end{table}

Table \ref{t6} illustrates message latency, hop count, and buffer time. Although the Epidemic protocol shows lower message latency than Spray and Wait, this is an illusion under low delivery rates. Fig.\ref{fig:11} shows the hop counts of information transmission under different protocols. Under the Epideimic protocol, the transmitted messages pass through a minimum of 1 hop and a maximum of 93 hops. Among them, 1 to 8 hops are relatively common, while cases exceeding 8 hops are less frequent. This indicates that the very few messages (75) delivered by Epidemic are almost all easily delivered or dropped without encountering buffer overflows. Messages requiring longer distances and difficult transmission are dropped when buffer overflows occur, with only a very small number of messages reaching delivery after dozens of hops. The Spray and Wait protocol, on the other hand, realistically demonstrates message forwarding latency in this use case, with an average of 1237.5179 seconds and a median of 800.6 seconds. All messages are delivered within the average number of hops, which lie between 1 and 5 hops, with an average of 3.2 hops. These buffer time statistics are very telling when it comes to the actual difference between the two types of protocols. Epidemic messages are backlogged for 18 seconds, but messages do not take too long to reach the target node within a short time and are easily deleted from the buffer, thus low buffer time, while Spray and Wait, because it can only replicate a limited number of times, has longer buffer times in rough seismic topologies, which causes messages to stay in the node buffer for a longer time and thus results in a larger average buffer time.

\begin{table}[htbp]
    \centering
    \caption{Message Transmit Comparison Between Epidemic and Spray and Wait in Nepal Earthquake Use Case}
    \begin{tabular}{ccc}
    \hline
        \textbf{Massage status} & \textbf{Epidemic} & \textbf{Spray and Wait} \\ \hline
        Latency Average & 405.8373 & 1,237.5179 \\ 
        Latency Median & 198.9 & 800.6 \\ 
        Hopcount Average & 10.6133 & 3.2620 \\ 
        Buffertime Average & 18.0119 & 5278.5112 \\ \hline
    \end{tabular}
    \label{t6}
\end{table}

\subsection{The Impact of Buffer Size on Communication Services}

In the Nepal Earthquake experiment, the rescue team nodes plays a significant role in message transmission. By modifying the buffer size of the rescue team nodes (10M, 50M, 100M), the performance of different protocols under different buffer sizes was observed. This analysis proved the superiority of the Spray and Wait protocol in the earthquake rescue use case.

\begin{figure}[htbp]
    \centering
    \includegraphics[width=1.0\linewidth]{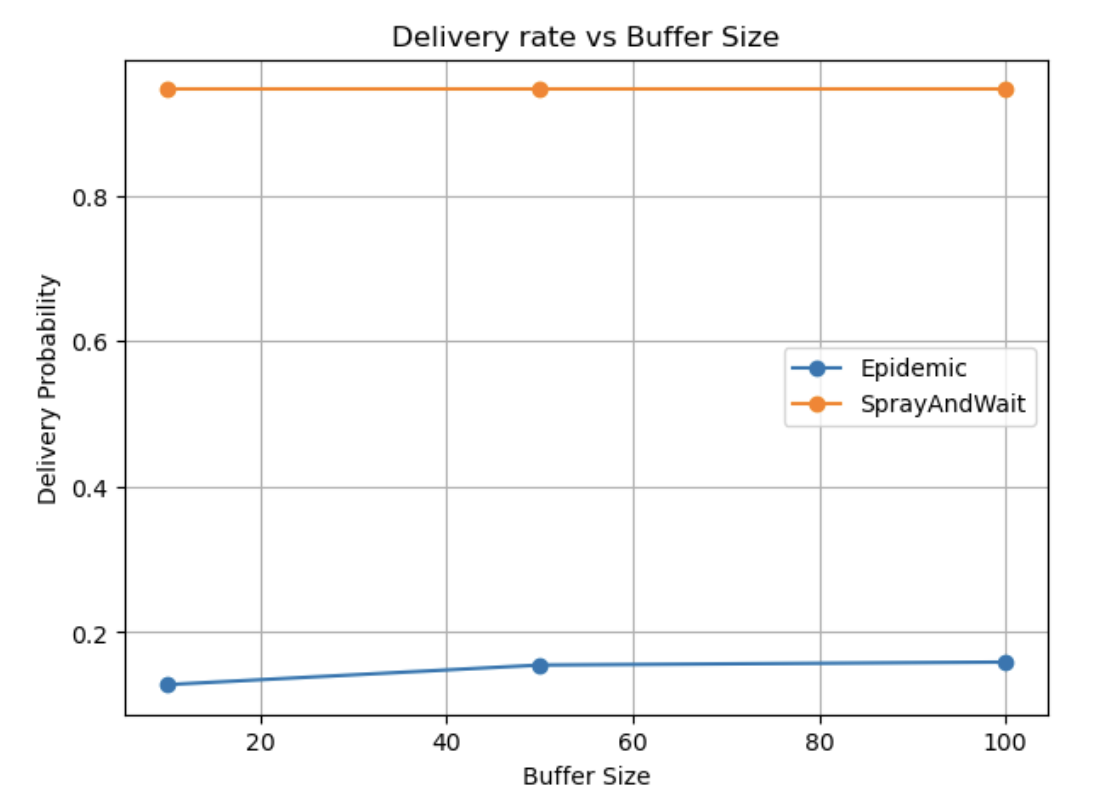}
    \caption{The delivery Probability of the Epidemic protocol and the Spray and Wait protocol under different buffer sizes in Nepal Earthquake Use Case}
    \label{fig:12}
\end{figure}

Tables \ref{t7} and \ref{t8}, and Fig.\ref{fig:12} shows the performance comparison in different protocols after adjusting the buffer. The delivery rate of the Epidemic routing protocol increases only slightly with the increase of buffer size. However, even with a buffer size as high as 100M, its delivery probability still cannot exceed 16\%. While the Spray and Wait protocol maintains a stable delivery rate of 94.6\% under all tested buffer sizes, proving its feasibility for deployment on resource-constrained small portable devices, which are typically used in emergency response. Since the Epidemic protocol adopts an unlimited replication strategy, the benefits brought by increasing the buffer size diminish; while the Spray and Wait protocol can effectively manage network resources and achieve better performance with less storage space.

\begin{table}[htbp]
\centering
\caption{The impact of buffer size on the Delivery probability of the Epidemic protocol in Nepal Earthquake Use Case}
\begin{tabular}{cccc}
\hline
\textbf{Massage status} &
\textbf{\makecell{Epidemic\\(10M)}} &
\textbf{\makecell{Epidemic\\(50M)}} &
\textbf{\makecell{Epidemic\\(100M)}} \\ 
\hline
Created & 484 & 484 & 484 \\ 
Delivered & 62 & 75 & 77 \\ 
Delivered Probability & 0.1281 & 0.1550 & 0.1591 \\ 
\hline
\end{tabular}
\label{t7}
\end{table}

\begin{table}[htbp]
\centering
\caption{The impact of buffer size on the Delivery probability of the Spray and Wait Protocol in Nepal Earthquake Use Case}
\begin{tabular}{cc}
\hline
\textbf{Massage status} &
\textbf{\makecell{Spray and Wait\\(10M / 50M / 100M)}} \\ 
\hline
Created & 484 \\ 
Delivered & 458 \\ 
Delivered Probability & 0.9463 \\ 
\hline
\end{tabular}
\label{t8}
\end{table}

\section{Conclusion and Future Work}
\subsection{Conclusion}
The experiment shows that Spray and Wait achieves better performance in disaster communications with limited resources than the conventional protocol, with a delivery success rate of more than 90\% and less network overhead. On the other hand, the Epidemic protocol crashes with a congestion collapse due to flood-based routing, with a delivery probability of no more than 20\%, plus tens of thousands of invalid transmissions. Seeing the protocol performance being degraded and not feasible for deployment, meanwhile, the hop count statistics show the stability of the Spray and Wait, with around five hops delivered for most messages. In comparison, the forwarding hop count of the Epidemic protocol in the scattered information states of several messages requires dozens of hops to deliver one or two messages. The buffer sensitivity analysis confirms that the congestion problem of the Epidemic can be solved only by increasing the storage space in limited degrees. Nevertheless, Spray and Wait, with a smaller buffer size of 10MB, also managed to retain its best performance. This implies that, in actual search and rescue operations, abilities applied in algorithms come to be more important than simply having big data storage hardware.

\subsection{Future work}
From the scenario perspective, the severe fire at Hongfu Estate in Tai Po, Hong Kong on November 26, 2025, caused at least 151 deaths and nearly 5,000 affected individuals\cite{b15}. Compared with earthquakes, fire scenarios may involve longer spatial evolution, different movement patterns, and dynamic network structures, including changing safe zones, reduced visibility, infrastructure collapse, and intermittent node failures. Evaluating DTN routing protocols under such conditions can improve understanding of robustness.

Improving DTN energy efficiency is another future direction. E3F optimizes energy consumption through packet prioritization, buffer management, and centrality-based forwarding decisions, maintaining high transmission performance with low energy usage\cite{b19}. Its adaptive storage and utility-driven forwarding mechanisms may extend network lifetime, particularly for assistance nodes and survivor devices with limited power. The SmartCharge algorithm introduces collaborative energy management using vehicular networking, edge computing, and Q-learning\cite{b20}, enabling nodes with sufficient power to support low-battery devices during rescue operations.

Future experimental validation may employ physical prototypes. The MODiToNeS platform provides a low-cost, lightweight framework for testing DTN algorithms in dynamic environments\cite{b21}. Its effectiveness has been demonstrated in agricultural monitoring and urban traffic monitoring scenarios using drones and vehicles\cite{b22}. With available connectivity, mobility, load, and power information, MODiToNeS can support earthquake case and evaluation of protocols such as Epidemic, Spray and Wait, congestion control, and cognitive caching.

\end{document}